# Indivisibility of Electron Bubbles in Helium

Veit Elser

*Department of Physics*
*Laboratory of Atomic and Solid State Physics*
*Cornell University*
*Ithaca, NY 14853-2501*

A recent proposal by Maris[1], that single electron bubbles in helium might fission into separate, particle-like entities, does not properly take into account the failure of the adiabatic approximation when, due to tunneling, there is a long electronic time scale. The point along the fission pathway of a photo-excited *p*-state bubble, where the adiabatic approximation first breaks down, occurs well before the bubble waist has pinched down forming two cavities. In the connected two-lobed geometry, the *p*- and *s*-states are strongly mixed by an antisymmetric vibrational mode, and the excitation decays by the mechanism where one lobe collapses while the other expands into the spherical *s*-state geometry. The extreme pressure jump in a photoexcited bubble leads to shock formation that may halt the elongation even before adiabaticity is compromised. In this case, the photoexcited bubble decays radiatively from the relaxed *p*-state geometry.[1]

---

1. In this manuscript the symbol $h$ in a mathematical expression denotes Planck's constant divided by $2\pi$.

# 1. Introduction

One of the most interesting laboratories for probing the general notion of "particle" is the bubble formed by a single electron injected into superfluid helium. The superfluid confers an enormous, essentially hydrodynamic, mass to the electron that is about 2 million times greater than its bare mass. At low enough temperatures, when bubble vibrations are frozen out and the fluid is free of excitations, the superfluid environment is in all respects a perfect, translationally and rotationally invariant vacuum through which the heavy electron propagates. Recently, Maris[1] has speculated that an even more exotic particle-like state of a single electron in helium is possible: an electron whose wavefunction has fissioned into two or more pieces, each of which supports a separate bubble. Moreover, he has suggested that the relaxation of photoexcited conventional electron bubbles might produce these exotic states and that fragmented bubbles might be the identities of previously unidentified species in mobility and photoconductivity measurements.

We argue here that the reasoning used to establish the theoretical possibility of fragmented electron bubbles lacked one important consideration which ultimately restores the indivisibility of the bubble state (Section 2). This is the failure to consider quantum fluctuations of the electron's environment, the bubble walls, in a regime where the manifold of electronic states has a comparable associated time scale. Such fluctuations are usually neglected in the description of the conventional bubble, where the scale of electronic excitations is large in comparison with the scale of the bubble's vibrational excitations. In the fragmented state, by contrast, there are long time scales associated with the tunneling of the electron between bubble fragments. It is then no longer valid to apply an adiabatic (Born-Oppenheimer) approximation to the bubble wall dynamics, and the correct treatment predicts unstable quantum fluctuations leading to the formation of a single bubble. We have calculated in some detail (Section 3) the exact moment in the proposed fissioning scenario where the collapse into a single bubble occurs.

Many features of photoexcitation experiments with electron bubbles are poorly understood and clearly hinge upon the nature of the relaxation process. In this connection (Section 4) we make the observation that the step-function change in the pressure on the bubble walls is so extreme as to generate velocities exceeding the Landau velocity by the time the bubble has elongated as little as 1Å in radius. This, together with the fact that the amplitude of the pressure jump exceeds the freezing pressure, suggests that the expansion may be highly overdamped even at zero temperature.

## 2. Breakdown of the adiabatic approximation

A useful device for developing an intuitive understanding of quantum many-body phenomena is the path integral representation of the partition function. As with the classical partition function, the probability of any particular configuration is derived from the Boltzmann weight for its energy relative to the thermal energy. In contrast to the classical partition function, however, a configuration for the fully quantum mechanical partition function includes the world lines of all the particles, with inverse temperature corresponding to the extent of the time axis. In isolation, each particle world line performs Brownian motion with a diffusion constant inversely proportional to the particle mass. If there is a repulsive interaction between particles, as in the system of helium atoms and one extra electron, the world lines will tend to avoid each other.

Figure 1(a) shows a typical configuration of world lines in a hypothetical world where the attractive part of the helium-helium potential is strong enough to favor a solid at low temperature. The helium solid has been prepared with two cavities which are preferred by the electron's world line in order to avoid the repulsive Pauli-exclusion interaction with the helium electrons. The quantum mechanical process of tunneling between cavities is represented in the path integral by configurations where the electron's world line jumps from one cavity to the other. Two important time/energy scales are represented by features in the electron world line in Figure 1(a): the fast traversal of a single cavity, $\tau_c$, and the much slower rate of jumping between cavities, $\tau_t$. These correspond to the large energy scale of levels within one cavity, $\Delta E_c = h/\tau_c$, in contrast to the smaller splitting, $\Delta E_t = h/\tau_t$, made possible by tunneling between cavities. As the cavities are moved apart, $\Delta E_t$ rapidly decreases while $\Delta E_c$ remains essentially unchanged.

Suppose we now remove the artificially enhanced attractive forces between helium atoms. The helium world lines will meander, although they still try to avoid each other and the electron. Since the mass of a helium atom is nearly four orders of magnitude larger than the electron mass, the helium world lines have root-mean-square velocities that are about 100 times smaller. This leads to a simplification in the analysis of the cavity/bubble occupied by the electron: the bubble walls can be treated quasi-statically. The diffusing electron world line performs many traversals of the cavity during a time when the bubble walls have moved a negligible amount. In the usual treatment of the bubble, this corresponds to imposing a fixed potential on the electron at the bubble walls whose position is determined by the balance between pressure from the electron's zero point motion and surface tension.

Clearly the equilibrium established between the electron and the helium world lines require the presence of the electron in the cavity. With no pressure being applied by the electron, the empty cavities in Figure 1(a) will collapse giving rise to the much more probable con-figuration shown in Figure 1(b). The collapse occurs when the root-mean-square helium velocity multiplied by $\tau_t$ is comparable with the radius of the cavity. Since $\tau_t$ increases with cavity separation, the collapse is inevitable. Alternatively we can sum this up with the statement that the adiabatic approximation fails precisely in the regime of the proposed multi-bubble state. A quantitative study of the breakdown of the adiabatic approximation, applied to a photoexcited bubble on the threshold of fissioning, is given in the next section.

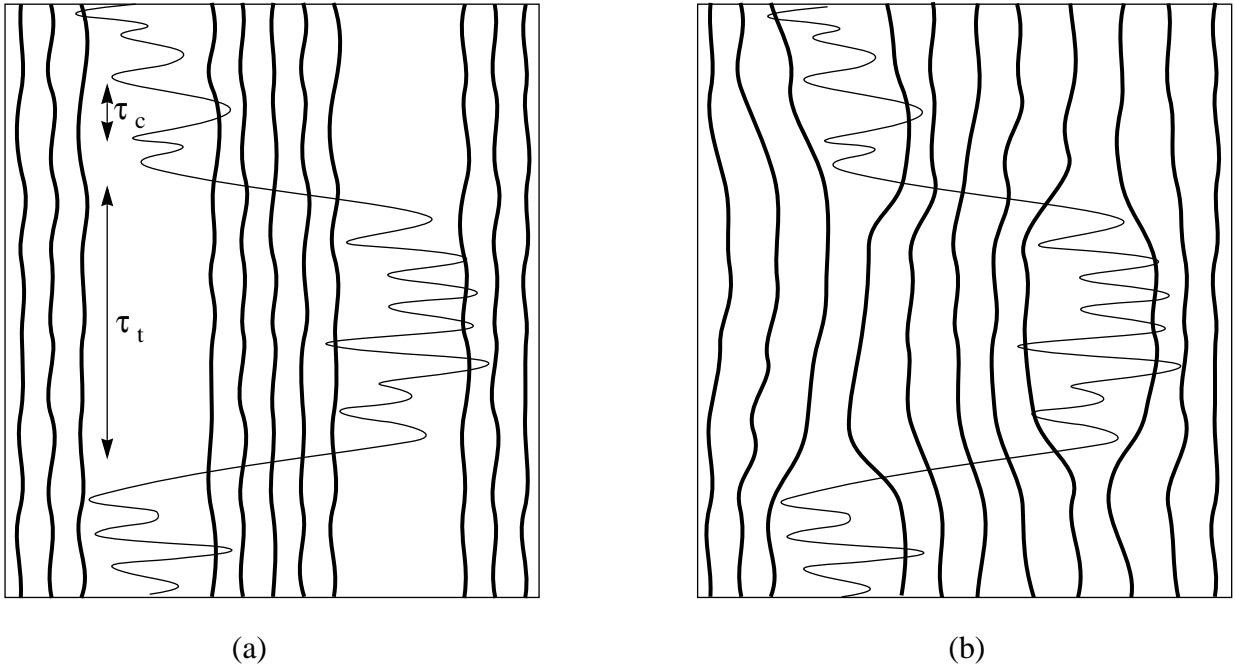

Figure 1. World lines in the path integral simulation of an electron (thin line) in helium (thick lines). (a) The electron tunnels between two cavities in a helium crystal. (b) In the liquid, cavities collapse unless the electron provides an outward pressure. The treatment of the cavities in the liquid in the manner of (a) represents an invalid application of the adiabatic approximation when there is a long electronic time scale ($\tau_t$).

What is the interpretation of Figure 1(b)? Not only the electron, but the entire bubble geometry is performing quantum fluctuations on a grand scale. These long-time/low-energy fluctuations are conceptually very straightforward, however: the quantum center-of-mass motion of a *single bubble*. The low energy dynamics of a single bubble is represented in the path integral by the diffusion of its center-of-mass. Because of its large (essentially hydrodynamic) mass, the world line of the bubble center will have an extremely small root-mean-square velocity. Consequently, the particular configuration shown in Figure 1(b) represents a relatively rare, but not forbidden, fluctuation in a single bubble's diffusive quantum motion.

## 3. Instability of the double-bubble

In the previous section we argued that since the tunneling time scale becomes arbitrarily long as the two cavities of a split bubble separate, there comes a point when this time scale matches the period of bubble vibrational oscillations and the adiabatic approximation fails. Here we show that this event occurs quite early, even before the bubble has completely fissioned. The geometry is shown in Figure 2: a single cavity formed from two identical spheres of radius $R_2$ joined along a circular orifice of radius $a$. This "double-bubble" is a reasonable model for the geometry encountered on the dynamical pathway of a fissioning bubble[1].

A ground ($1s$) state spherical bubble photoexcited to the first excited ($1p$) state exhibits the Jahn-Teller effect and responds by expanding uniaxially, the planar node of the electronic wavefunction bisecting the prolate bubble into reflection symmetric halves. Since the pressure of the confined electron vanishes at the node, the waist of the bubble will shrink even while the two poles of the bubble expand. In the double-bubble geometry, right on the threshold of fission, the node of the electronic wavefunction spans the orifice so that the two halves approximate simple spherical bubbles with a vanishing wavefunction over their entire surface. The double-bubble is an equilibrium shape only in the limit $a \to 0$; for finite $a$ the equilibrium geometry has a smooth waist and lower energy than can be achieved by the double-bubble.

In determining energies and shapes of bubbles we include the electron kinetic energy and the helium surface energy, where the former is calculated in the approximation which neglects wavefunction penetration into the helium. We also neglect the weak polarization of the helium by the electron's electric field and confine our attention to the case of zero applied pressure. With these approximations, energies may be expressed in terms of a length formed from the

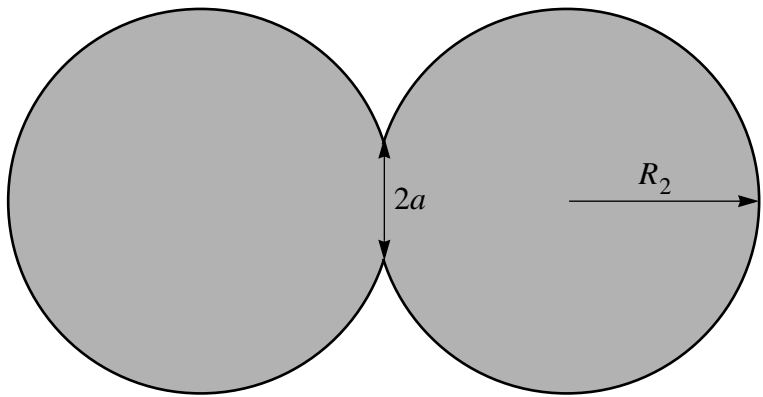

Figure 2. Double-bubble geometry

## Table 1: Sizes and energies of bubble geometries

| shape/state | radius ($R_0$) | radius (Å) | energy ($E_0$) | energy (eV) |
|---|---|---|---|---|
| spherical bubble | $R_1 = \left(\dfrac{\pi}{4}\right)^{1/4}$ | $R_1 = 19.3$ | $E_1 = 4\pi^{3/2}$ | $E_1 = 0.200$ |
| photoexcited spherical bubble[a] | $R_1$ | $R_1$ | $E_1^* = 33.88$ | $E_1^* = 0.304$ |
| relaxed double-bubble ($a = 0$) | $R_2 = \left(\dfrac{\pi}{8}\right)^{1/4}$ | $R_2 = 16.2$ | $E_2 = 2\sqrt{8\pi^3}$ | $E_2 = 0.283$ |
| double-bubble with energy $E_1^*$ [a] | $R_2^* = 0.960$ | $R_2^* = 19.7$ | $E_1^*$ | $E_1^*$ |
| relaxed $p$-state bubble[b] | $R_p = 1.41$ | $R_p = 29.0$ | $E_p = 29.6$ | $E_p = 0.266$ |

a. Requires numerical zero of spherical Bessel function $j_1$.
b. Numerical solution for general, axially symmetric, geometry; $R_p$ is the semi-major axis[1,2].

electron mass $m_e$ and the helium surface tension $\sigma$:

$$R_0 = \left(\frac{\hbar^2}{2m_e\sigma}\right)^{1/4} = 20.5 \text{ Å}. \quad (1)$$

The basic unit of energy is given by:

$$E_0 = \sigma R_0^2 = \frac{\hbar^2}{2m_e R_0^2} = 8.98 \text{ meV}. \quad (2)$$

A collection of lengths and energies of various static bubble geometries is given in Table 1.

The energies of the double-bubble in Table 1 are calculated in the limit of vanishing orifice radius ($a = 0$), so that the symmetric ($1s$) and antisymmetric ($1p$) electronic states are exactly degenerate in the limit of zero helium penetration. More realistically, the energies of these states are split by tunneling of the electron through the thin helium barrier. The corresponding time scale, when compared with the period of bubble vibrations, can be used to pinpoint the breakdown of the adiabatic approximation. On the other hand, a long enough time scale already occurs without the need for tunneling, when the electron is able to pass through a finite orifice. The energy splitting $\Delta = E_{1p} - E_{1s}$, in the limit of impenetrable helium, can be cal-culated perturbatively for $a \ll R_2$; the leading term is (Appendix A):

$$\Delta = \frac{4\pi}{3} E_0 \left(\frac{R_0}{R_2}\right)^2 \left(\frac{a}{R_2}\right)^3. \tag{3}$$

An electron in a superposition of $1s$ and $1p$ states in the double-bubble has a probability density that oscillates between the two halves with frequency $\Delta/h$. The pressures on the two spherical walls will therefore have an antisymmetric component that oscillates with the same frequency and drives the vibrational mode shown in Figure 3. We use $r$ as the dynamical variable for this mode, where the two halves of the double-bubble have radii $R_A = R_2 + r$ and $R_B = R_2 - r$. There is a qualitative similarity between this antisymmetric breathing mode and the mode responsible for the collapse of the fissioned bubble in Figure 1.

The stiffness of the antisymmetric breathing mode, in the adiabatic approximation, has contributions from the electron kinetic energy and the surface energy:

$$U = E_\pm + 4\pi\sigma(R_A^2 + R_B^2), \tag{4}$$

where $E_\pm$ are the two eigenvalues of the two-state system of an electron in coupled spherical cavities of radii $R_A$ and $R_B$ [1]:

$$E_\pm = \frac{E_A + E_B \pm \sqrt{(E_A - E_B)^2 + \Delta^2}}{2}. \tag{5}$$

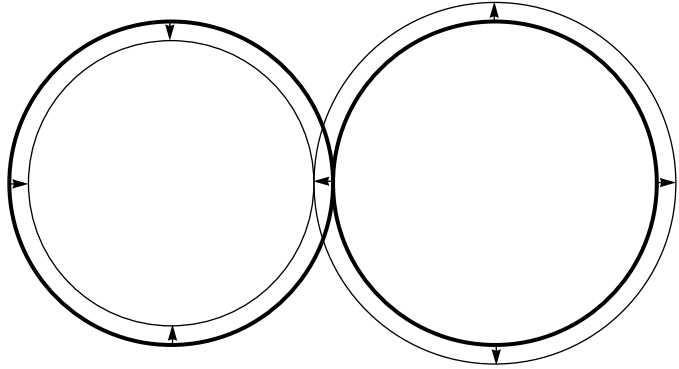

Figure 3. Antisymmetric breathing mode.

The coupling of the cavities has been chosen so that for $E_A = E_B$, the two eigenvalues are split by $\Delta$, while $E_+$ and $E_-$ correspond to $E_{1p}$ and $E_{1s}$ respectively. Expanding $U$ to lowest order in $r$ we obtain:

$$\delta U = \pm 8 \frac{E_0^2}{\Delta}\left(\frac{r}{R_2}\right)^2 + 8\pi\sigma r^2. \tag{6}$$

In the limit $\Delta \to 0$ (and for the finite values considered below) the electronic term dominates the surface term which henceforth will be neglected. As noted by Maris[1], only the $1p$ state ($E_+$) has a positive stiffness and supports vibrational oscillations.

The other ingredient in the determination of the antisymmetric breathing frequency is the inertia of the mode. For this we treat the superfluid as an inviscid, incompressible continuum fluid whose velocity field is uniquely determined by a solution of Laplace's equation with appropriate boundary conditions. For the mode in question we impose Neumann boundary conditions, specifically, with constant normal derivatives equal to $\pm\dot{r}$ on the two spheres. The resulting superfluid kinetic energy, in the limit of two touching spheres ($a = 0$), is given by (Appendix B):

$$K = 2\pi\mu\rho R_2^3 \dot{r}^2, \tag{7}$$

where $\rho$ is the helium mass density and $\mu \cong 1.70168\ldots$ is a numerical constant whose value would be exactly 2 in the limit of infinitely separated bubble halves.

Combining the stiffness with the inertia of the oscillator we obtain the frequency of the antisymmetric breathing mode:

$$\omega_- = \sqrt{\left(\frac{4}{\pi\mu}\right)\frac{E_0^2}{\rho R_2^5 \Delta}}. \tag{8}$$

The adiabatic approximation breaks down when $\hbar\omega_- = \Delta$, or equivalently,

$$\frac{a}{R} = \left(\frac{27}{8\mu}\right)^{1/9}\left(\frac{m_e}{\rho R_0^3}\right)^{1/9}\left(\frac{R_0}{R_2}\right)^{1/3}$$
$$= (0.22)\left(\frac{R_0}{R_2}\right)^{1/3}. \tag{9}$$

From Table 1 we see that $R_2$ ranges from about 16 Å for a fully relaxed double-bubble, to about 20 Å, if all the energy of photoexcitation is available (no dissipation). The orifice diameter at the breakdown of the adiabatic approximation thus ranges from about 7.5 to 9 Å.

We conclude from this analysis that the $1p$ state becomes thoroughly mixed with the $1s$ state at a time when the waist of the expanding photoexcited bubble is still relatively large, about 8 Å in diameter. The result of this mixing is that the stabilizing stiffness of the $1p$ state erodes and is replaced by the destabilizing effect of the $1s$ state. This can be interpreted as a non-radiative deexcitation of the bubble, where a small quantum of electronic energy $(E_{1p} - E_{1s})$ is transferred to a vibrational mode ($h\omega_-$). While in the $1s$ state, the antisymmetric breathing mode is unstable: one half expands into a spherical $s$-state bubble while the other collapses and vanishes completely. This scenario was observed in simulations of an excess electron in a classical helium fluid ($T = 309$ K) by Space and Coker[3]. The collapse of the double-bubble is a violent event and, using the numbers in Table 1, is expected to release 80-100 meV, mostly in the form of rotons and small vortex rings.

## 4. Extreme non-equilibrium conditions in a photoexcited bubble

It is quite likely that upon photoexcitation the *p*-state bubble never reaches the pre-fission double-bubble geometry (Section 3) because of strong damping associated with the high stresses imposed by the excited electron. Immediately after photoexcitation to the 1*p*, *m* = 0 state, the pressure exerted on the spherical bubble walls,

$$P_{1p} = \frac{h^2}{2m_e}|\nabla \Psi_{1p}|^2 \tag{10}$$

has monopole as well as quadrupole terms. The pressure jump, relative to the pressure in the equilibrium *s*-state bubble, is given by

$$\Delta P = P_{1p} - P_{1s} = \frac{E_0}{R_1^3}\left(\frac{R_0}{R_1}\right)^2 \left[\left(\frac{\alpha}{3} - \frac{\pi}{2}\right) + \frac{2\alpha}{3}P_2(\cos\theta)\right], \tag{11}$$

where $P_2$ is the second Legendre polynomial, $\theta$ is the polar angle on the bubble, and $\alpha = x_1^5[j_1'(x_1)]^2 = 9.64...$ is a numerical constant involving the first zero ($x_1$) of the spherical Bessel function $j_1$. The maximum pressure occurs at the poles:

$$\Delta P_{max} = (13.3 \text{ atm})\left(\frac{R_0}{R_1}\right)^5. \tag{12}$$

This shows that for $R_1 < 18$ Å, the pressure change upon photoexcitation exceeds the freezing pressure of helium. At elevated ambient pressures, where a smaller pressure change induces freezing and $R_1$ is also smaller, the potential for the bubble to locally solidify the helium is even greater.

The stress applied to the helium by the photoexcited electron is far from hydrostatic and the first response will be for the helium to be accelerated as a fluid. The behavior for short times after the excitation can be modeled using just the monopole and quadrupole modes of the bubble shape:

$$R(\theta, t) = R_1 + a_0(t) + a_2(t)P_2(\cos\theta). \tag{13}$$

Assuming, at least for very early times, that the helium flows as an incompressible superfluid,

the corresponding kinetic energy in the flow is given by

$$K = 2\pi\rho R_1^3\left(\dot{a}_0^2 + \frac{\dot{a}_2^2}{15}\right). \tag{14}$$

This kinetic energy is generated by the work performed by the electron pressure on the superfluid:

$$\begin{aligned}W &= \int \Delta P(r-R_1)2\pi R_1^2 d(\cos\theta) \\ &= 4\pi\left(\frac{E_0}{R_0}\right)\left(\frac{R_0}{R_1}\right)^3\left[\left(\frac{\alpha}{3}-\frac{\pi}{2}\right)a_0 + \frac{2\alpha}{15}a_2\right]\end{aligned} \tag{15}$$

From $K$ and $W$ one obtains the accelerations of the two mode amplitudes:

$$\begin{aligned}\ddot{a}_0 &= g_0 = \left(\frac{\alpha}{3}-\frac{\pi}{2}\right)\left(\frac{E_0}{R_0^4\rho}\right)\left(\frac{R_0}{R_1}\right)^6 \\ \ddot{a}_2 &= g_2 = 2\alpha\left(\frac{E_0}{R_0^4\rho}\right)\left(\frac{R_0}{R_1}\right)^6\end{aligned}. \tag{16}$$

The most extreme accelerations and highest velocities occur at the poles; using $z = a_0 + a_2$ to represent the displacement of the bubble wall there, we obtain:

$$\begin{aligned}\dot{z} &= \sqrt{2(g_0+g_2)z} \\ &= (48 \text{ m/s})\left(\frac{R_0}{R_1}\right)^3\sqrt{\frac{z}{\text{Å}}}\end{aligned} \tag{17}$$

The calculation above shows that by the time the bubble has elongated about 1Å, the fluid velocity already exceeds the Landau critical velocity for roton creation. Under these conditions one expects a shock to form which may take the form of a large pressure gradient. Thus it is conceivable that almost immediately after photoexcitation, a highly compressed, perhaps even solid, region develops in the fluid that is supported by the electron pressure on one side and the pressure drop in the shock on the other. This will halt the expansion much more effectively than if all the work performed by the expanding electron wavefunction is transformed into fluid kinetic energy. Eventually, when a new equilibrium can be established between the electron pressure and the Laplace pressure of the curved bubble wall, the acceleration ceases and the region of shocked, highly compressed helium expands. The expansion of the shocked helium into the surrounding liquid should be a very efficient means of thermalizing the excess energy of the photoexcited bubble ($E_1^*$) relative to the energy of the relaxed *p*-state bubble ($E_p$).

If the expansion of the photoexcited electron bubble is strongly damped, as the calculation above suggests, then it is quite possible that elongation will stop short of the double-bubble geometry where, because of non-adiabaticity, a non-radiative relaxation mode is available. In the strong damping scenario, the electronic wavefunction would remain adiabatically *p*-like, eventually reaching the equilibrium *p*-state geometry when the excess energy has been radiated away in the form of rotons and phonons. The equilibrium *p*-state bubble eventually decays radiatively to the *s*-state, which then relaxes to the spherical geometry.

# 5. Conclusions

The strangeness of a quantum state is bounded only by the powers of the imagination; a more relevant consideration is its likelihood of being observed. Thus it is one thing to "prepare" a system in a particular state with interesting properties, quite another to expect the system to retain these properties over time. In the helium-electron system one can construct a state, such as rendered in Figure 1(a), where an electron tunnels between two relatively static cavities in the fluid. This state squanders a fair amount of what might be called "correlation energy": an energy reducing correlation in the breathing mode of, and the electron's residency within, the two cavities. A better representation of the ground state, which exploits this correlated motion, is Figure 1(b): a single bubble in a superposition of two definite position states. The "split bubble", Fig. 1(a), is really a vibrational excited state relative to the single bubble, Fig. 1(b), and as such will quickly radiate its excess energy in the form of phonons and rotons.

Even though the split bubble is not a viable, near equilibrium state, Maris' question[1] regarding the final state in photoexcitation experiments is an interesting one. We have considered two scenarios, depending on the strength of the damping experienced by the expanding bubble. If the damping is not too strong, and the elongation of the photoexcited $p$-state bubble is able to reach a point where its waist has contracted to a diameter of about 8Å, then a nonradiative decay of the excitation will occur. In the near-fission geometry of the "double-bubble", the $1p$-$1s$ electronic splitting matches the frequency of antisymmetric breathing oscillations of the bubble walls. This mixes the electronic states and the destabilizing influence of the $1s$ state on the breathing mode leads to the eventual collapse of one bubble half. In view of the exact reflection symmetry of the intial photoexcited state, the asymmetry of this nonadiabatic mechanism implies the final spherical bubble will be in a superposition of two position states, one corresponding to each half of the parent double-bubble. There is nothing about this particular superposition, however, that is different in principle from other forms of superposition (e.g. wavepackets) in the description of single particle states.

The decay of photoexcited bubbles by the nonadiabatic mechanism is likely to be violent, a point of comparison being the collapse of cavitation bubbles in a classical fluid near walls[4]. Self-focussing effects in the latter leads to the formation of a sharp jet on one side of the bubble that then penetrates the opposite side. In a collapsing electron double-bubble, the "jet" formed from the collapse of one bubble half may be as concentrated as just a few very energetic helium atoms that are launched with high velocity through the surviving spherical bubble half, reentering the fluid as a highly directional shower of rotons[5].

In the other decay scenario, the expansion is halted by shock formation due to the extreme pressure jump on the photoexcited bubble. The magnitude of the pressure jump exceeds the freezing pressure of helium and, in the absence of dissipation, will quickly generate bubble wall velocities in excess of the Landau velocity. These conditions suggest that the bubble expansion creates a region of highly compressed helium that first stores, and eventually releases, the work performed by the expanding electron wavefunction. If the dissipation is sufficient, the bubble expansion will fall short of reaching the necessary slim waistline where the non-adiabatic 1$p$ to 1$s$ conversion of the electronic state is possible. Eventually the $p$-state bubble will reach the equilibrium $p$-state geometry and decay radiatively. The light from this luminescent decay mode would have the striking experimental signature of being redshifted relative to the original photoexcitation by over 30%.

Evidence from photoconductivity experiments by Grimes and Adams[6] can be construed to support a pressure-dependent competition between the two scenarios. Inferences based on these experiments have to be made cautiously, however, since the details of the transport mechanism are largely unknown. There is good reason to believe that the electron bubbles are trapped on vortex lines and that the conductivity signal associated with photoexcitation involves a transition to the untrapped state. The characteristic of the two scenarios that could have a bearing on this transport speculation is the symmetry of the deexcitation process. In the non-adiabatic process, the one-sided collapse of the double-bubble is very asymmetrical and a significant momentum can be transferred to the final spherical bubble, possibly enough to eject it from a vortex line. On the other hand, in the overdamped expansion scenario, the bubble maintains reflection symmetry throughout the relaxation to the equilibrium $p$-state geometry. Since the bubble does not acquire a significant momentum, it remains trapped and there is no photoconductivity signal. The vanishing of the photoconductivity signal below 1 atm, observed by Grimes and Adams[6], can thus be interpreted as a suppression of the nonadiabatic decay mode at low pressures. Observation of a luminescence signal, correlated with the decrease in the photoconductivity signal, would be strong evidence in favor of this interpretation.

# 7. Appendix A: Tunnel splitting in the double-bubble

The approximate energy splitting between the reflection-symmetric ground state ($1s$) and the antisymmetric first excited state ($1p$) of the symmetric double-bubble can be obtained using a technique originally developed to find frequency shifts in waveguides[7]. The first step is to replace the double-bubble geometry by a single sphere with an orifice, and identify the two states as ground states of the single sphere having different boundary conditions on the orifice. Since we are interested in the case of a small orifice, we may also replace the planar orifice of the original reflection-symmetric problem with a small spherical cap. With this approximation, the $1p$ wavefunction becomes the ground state in a sphere with zero Dirichlet boundary condition over the entire surface:

$$\Psi_{1p} = j_0\left(\pi \frac{r}{R}\right)$$

$$E_{1p} = \pi^2 \frac{h^2}{2m_e R^2}$$

(18)

The $1s$ state differs from the $1p$ state by having a zero Neumann boundary condition over the orifice. Since the orifice is small, we express the $1s$ wavefunction perturbatively as

$$\Psi_{1s} = \Psi_{1p} + \delta\Psi,$$

(19)

and expand the expectation of the kinetic energy to first order in $\delta\Psi$:

$$\begin{aligned}
E_{1s} &= \frac{h^2}{2m_e} \frac{\int_V |\nabla \Psi_{1s}|^2}{\int_V |\Psi_{1s}|^2} \\
&\cong E_{1p} + \frac{h^2}{m_e} \frac{\int_S (\hat{n} \cdot \nabla \Psi_{1p}) \delta\Psi}{\int_V |\Psi_{1p}|^2}
\end{aligned}$$

(20)

($V$ represents the volume of the sphere, $S$ its surface, and $\hat{n}$ is the outward surface normal). Since $\delta\Psi$ vanishes everywhere on $S$ with the exception of the orifice, we obtain:

$$\Delta \cong \frac{h^2}{m_e} \frac{\int_{\text{orifice}} (\hat{n} \cdot \nabla \Psi_{1p}) \delta\Psi}{\int_V |\Psi_{1p}|^2}$$

(21)

To obtain the short length scale behavior of $\Psi_{1s}$ in the vicinity of the orifice, it suffices to solve the corresponding electrostatics problem, $\nabla^2 \Psi_{1s} \cong 0$, for a planar grounded conductor with a circular hole of radius $a$. The boundary condition for this problem is the asymptotic "electric field" on the two sides of the conductor which we take to be the gradient of the unperturbed ($1p$) wavefunction on the surface of the sphere. An explicit formula for the electrostatic potential in the plane of the hole is given in Jackson[7]:

$$\delta\Psi = \frac{2}{\pi} G \sqrt{a^2 - \rho^2}$$
$$G = -\hat{n} \cdot \nabla \Psi_{1p}$$
(22)

($\rho$ is the distance from the center of the hole). Using this result in the expression for $\Delta$ above, we obtain:

$$\Delta = \frac{2\pi}{3}\left(\frac{h^2}{m_e R^2}\right)\left(\frac{a}{R}\right)^3.$$
(23)

# 8. Appendix B: Kinetic energy of the antisymmetric breathing mode

In the continuum approximation, where the helium is treated as an inviscid, incompressible fluid, the kinetic energy in the flow is given by

$$K = \frac{1}{2}\rho \int_V |\nabla \Phi|^2, \tag{24}$$

where $\rho$ is the helium mass density, $V$ is the volume occupied by fluid, and the velocity potential satisfies $\nabla^2 \Phi = 0$. The boundary conditions on $\Phi$ emerge more explicitly when integration by parts is used to reexpress $K$ as

$$K = \frac{1}{2}\rho \int_S (\hat{n} \cdot \nabla \Phi) \Phi \tag{25}$$

where $S$ is the surface of the bubble and $\hat{n}$ is the surface normal pointing into the bubble. For the antisymmetric breathing mode $\hat{n} \cdot \nabla \Phi$ is constant and equal to $\pm \dot{r}$ on the two spherical halves; consequently,

$$K = \rho \dot{r} \int_{S_+} \Phi, \tag{26}$$

where $S_+$ denotes the sphere with radial velocity $+\dot{r}$.

The velocity potential is obtained using the method of images applied to a dipole distribution of sources. It is well known that a pair of monopole sources solves the zero-Dirichlet boundary value problem on a sphere[7]. For the zero-Neumann boundary condition, corresponding to vanishing flow across the sphere, a pair of dipole sources provides the basis for a solution by images[8]. The basic dipole pair, for a sphere of radius $R$ centered at the origin, is the source dipole of unit moment $\hat{r}$ at position $r$, and the image with moment $-(R/r)^3 \hat{r}$ located at $(R/r)^2 r$. In constructing the solution for the double bubble, we will be interested in a distribution of dipoles on the axis of symmetry. If the latter is the $z$-axis, and the two spheres of the double-bubble are located at $z = \pm R$, then the corresponding images of a linear dipole density $q(z)$ are the densities

$$\begin{aligned} q_+(z) &= -\frac{|z-R|}{R} q\left(\frac{R^2}{z-R} + R\right) \\ q_-(z) &= -\frac{|z+R|}{R} q\left(\frac{R^2}{z+R} - R\right) \end{aligned} \tag{27}$$

upon inversion in the spheres at +R and -R respectively.

For the initial dipole distribution we use the step function

$$q_0(z) = \begin{cases} -R^2 \dot{r} & z > R \\ 0 & z < R \end{cases} \quad (28)$$

which is equivalent to a monopole source of strength $R^2 \dot{r}$ located at $z = R$, corresponding to a velocity field $-\nabla \Phi = \dot{r}\hat{r}$ on the surface of the sphere at $z = R$. To satisfy the zero-normal-velocity boundary condition on the sphere at $z = -R$, we invert $q_0$ using the formula for $q_-$ to obtain $q_1$. This in turn must be inverted, using $q_+$, to obtain $q_2$, etc. The semi-infinite sequence of images obtained this way is shown in Figure 4. Reflecting the entire sequence about $z = 0$ gives us the sources required to satisfy the boundary condition $-\nabla \Phi = -\dot{r}\hat{r}$ on the other sphere, at $z = -R$. The superposition of both sequences solves the original problem; the entire dipole density is given by the formula

$$q(z) = \begin{cases} -R^2 \dot{r} & R < z \\ -(-1)^n (R^2 - nRz)\dot{r} & \dfrac{R}{n+1} < z < \dfrac{R}{n} \quad n = 1, 2, \ldots \\ q(-z) & z < 0 \end{cases} \quad (29)$$

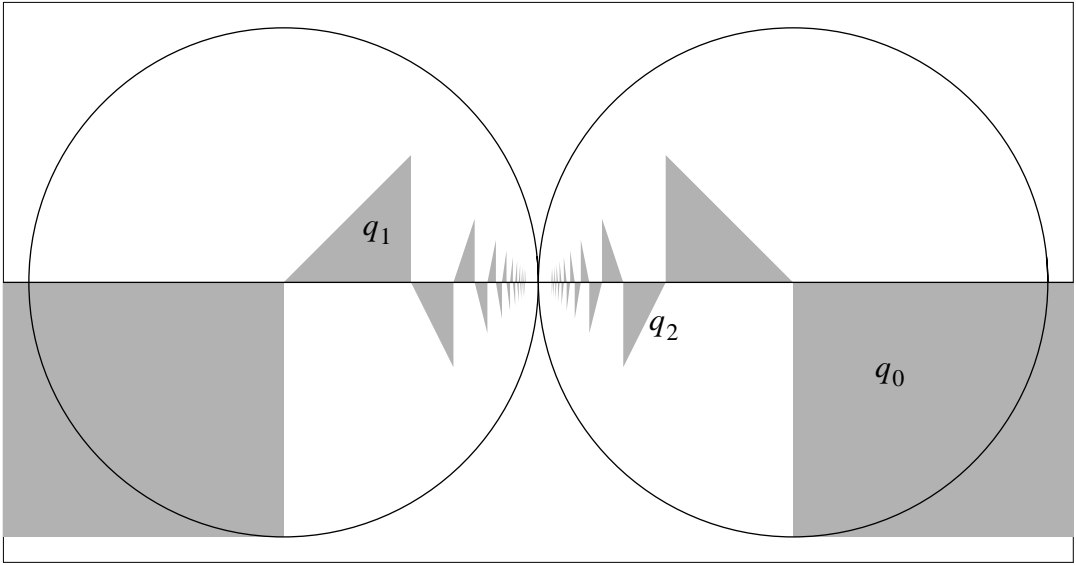

Figure 4. Dipole source distribution (gray) along the axis of symmetry of the double-bubble for the flow field of the antisymmetric breathing mode (Fig. 3). The sequence of image distributions $q_1, q_2, \ldots$ is formed by inverting the initial distribution $q_0$ in alternating spheres.

The negative derivative, $-q'(z)$, gives the equivalent monopole source density, for which we can use the expansion of the free Green function appropriate for the surface of the sphere at $z = R$:

$$G_+(\theta;z) = \begin{cases} \sum_{m=0}^{\infty} \frac{|z-R|^m}{R^{m+1}} P_m(\cos\theta) & |z-R| < R \\ \sum_{m=0}^{\infty} \frac{R^m}{|z-R|^{m+1}} P_m(\cos\theta) & |z-R| > R \end{cases} \qquad (30)$$

Integrating $G_+$ over the sphere and source distribution we obtain the kinetic energy:

$$K = \rho\dot{r}(4\pi R^2)\left[\frac{q(0)-q(2R)}{R} - \int_{-\infty}^{0} \frac{q'(z)}{|z-R|} dz\right] \qquad (31)$$

$$= 2\pi\mu\rho R^3 \dot{r}^2$$

The boundary term above, due to sources inside the sphere, just gives the coefficient value $\mu = 2$ corresponding to a pair of infinitely separated spheres. When combined with the integral, the kinetic energy coefficient has the value

$$\mu = \log 4 + 2 \sum_{n=1}^{\infty} (-1)^n n \log\left[1 - \frac{1}{(n+1)^2}\right] \qquad (32)$$

$$\cong 1.70168$$